\documentclass[12pt]{iopart}
\usepackage[applemac]{inputenc}
\usepackage{iopams}
\usepackage{graphicx}

\begin{document}

\title[Implementing two-photon interference in the frequency domain]{Implementing two-photon interference in the frequency domain with electro-optic phase modulators}

\author{Laurent Olislager$^1$, Isma\"el Mbodji$^2$, Erik Woodhead$^3$, Johann Cussey$^2$, Luca Furfaro$^2$, Philippe Emplit$^1$, Serge Massar$^3$, Kien Phan~Huy$^2$, and Jean-Marc Merolla$^2$}
\address{$^1$Service OPERA-Photonique, CP 194/5, Universit\'e Libre de Bruxelles, Avenue F. D. Roosevelt 50, B-1050~Brussels, Belgium}
\address{$^2$D\'epartement d'Optique P.M. Duffieux, Institut FEMTO--ST, Centre National de la Recherche Scientifique, Unit\'e Mixte de Recherche 6174, Universit\'e de Franche-Comt\'e, Route de Gray 16, F-25030 Besan\c{c}on, France}
\address{$^3$Laboratoire d'Information Quantique, CP 225, Universit\'e Libre de Bruxelles, Boulevard du Triomphe, B-1050 Brussels, Belgium}
\eads{\mailto{lolislag@ulb.ac.be}, \mailto{ismael.mbodji@univ-fcomte.fr}}
\begin{abstract}
Frequency-entangled photons can be readily produced using parametric down-conversion. We have recently shown how such entanglement could be manipulated and measured using electro-optic phase modulators and narrow-band frequency filters, thereby leading to two-photon interference patterns in the frequency domain. Here we introduce new theoretical and experimental developments showing that this method is potentially a competitive platform for the realization of quantum communication protocols in standard telecommunication fibres. We derive a simple theoretical expression for the coincidence probabilities and use it to optimize a Bell inequality. Furthermore, we establish an equivalence between the entangled-photon scheme and a classical interference scheme. Our measurements of two-photon interference in the frequency domain yield raw visibilities in excess of 99\%. We use our high quality setup to experimentally validate the theoretical predictions, and in particular we report a violation of the CH74 inequality by more than 18 standard deviations.
\end{abstract}
\maketitle

\section{Introduction \label{sec:introduction}}

Precision manipulation of entangled photons is highly desirable, both from the fundamental point of view of studying the ultimate limits of optics and from the point of view of applications such as quantum communication. Indeed, since Ekert's seminal work \cite{ref:e91}, entangled photons have appeared to be a promising way to distribute quantum information. Using entangled photons could potentially allow the realization of key distribution protocols over distances greater than a few hundred kilometres \cite{ref:wzy02,ref:mfl07,ref:sst11} and security certification without \textit{a priori} trust in the devices employed \cite{ref:ab07}.

Most practical quantum key distribution methods based on entangled photons use time-bin \cite{ref:er92,ref:tb00} or polarization \cite{ref:js00,ref:np00} encoding. These have also been among the preferred methods for investigating the fundamental issue of quantum nonlocality \cite{ref:agr82,ref:f89,ref:tb99}. Manipulating entangled photons directly in the frequency domain is a relatively unexplored area. Previous work in this direction includes Hong--Ou--Mandel dip experiments \cite{ref:om88,ref:kkk03,ref:fh09}, creation of entanglement in multiple degrees of freedom including frequency \cite{ref:ss94,ref:bl05} and conversion from polarization to frequency entanglement \cite{ref:rr09}.

In \cite{ref:oc10}, we introduced the notion of \emph{frequency-bin entanglement} that allows a simple description of experiments that manipulate entanglement in the frequency domain. We have shown how, using conventional methods of production (parametric down-conversion) and detection (avalanche photodiodes (APDs)), frequency-bin entangled photons at telecommunication wavelengths (about 1550~nm) could be manipulated in optical fibres using standard telecommunication components such as fibre Bragg gratings and electro-optic phase modulators (EOPMs) driven by radio-frequency (RF) signals.

In this work, we improve on the work reported in \cite{ref:oc10}. We develop the theory behind this experiment as well as report experimental improvements. The results reported in \cite{ref:oc10} and in the present paper build on earlier experimental investigations on quantum communication using attenuated coherent states and frequency encoding \cite{ref:mm99,ref:bm07}, which in particular aimed at applications in quantum key distribution. Complementary theoretical studies of the manipulation of photons in the frequency domain using EOPMs can be found in \cite{ref:cf10,ref:cf11}.

The first part of the paper deals with the theoretical aspects of our experiment involving frequency-bin entangled photons and EOPMs. After describing the basic scheme, we show that it is possible to derive a considerably simpler expression for the joint probabilities than was reported in \cite{ref:oc10}. This shows that, whereas in most experiments involving entangled photons the interference pattern is given by sine functions, here it is given by Bessel functions. We then use this expression to derive the optimal RF settings for violation of a Bell-type inequality, the CH74 inequality. Finally, we adapt the correspondence between “prepare-and-measure" and entanglement-based schemes (often exploited in quantum key distribution) to our setup, showing that identical RF settings will give rise to identical interference patterns both in two-photon and in single-photon experiments. This identity is very useful experimentally as it makes it possible to test the quality of the RF setup using a broadband white light source. A detailed comparison of the different experiments is reported in section \ref{sec:setup}.

At the experimental level, we have improved the experiment reported in \cite{ref:oc10} in a number of ways. We have developed an RF architecture based on off-the-shelf components that provides highly stable control of both the amplitude and phase of the RF signals used to drive the EOPMs. In addition, we have improved the stability of the pump laser wavelength and optimized the conversion efficiency of the periodically poled lithium niobate (PPLN) crystal. We now also use low-noise superconducting single-photon detectors (SSPDs). Altogether, these improvements allow us to report two-photon interference with raw and net visibilities of $(99.17\pm0.11)\%$ and $(99.76\pm0.11)\%$, respectively. These are comparable to the best results reported for two-photon interference at telecommunication wavelengths; see, e.g., \cite{ref:mi10}. By using the optimal settings mentioned above, we also report violation of the CH74 inequality by more than 18 standard deviations. Both these figures are considerably better than those reported in \cite{ref:oc10}.

These results lay the groundwork for future experiments. In particular, they show that manipulating frequency-bin photon entanglement with EOPMs is a promising platform for the realization of quantum communication protocols at telecommunication wavelengths.

\section{Theoretical analysis\label{sec:intuition}}

\subsection{Manipulating and measuring frequency-entangled photons using electro-optic phase modulators and narrow-band frequency filters}

The experimental scheme we study in this paper is depicted in figure~\ref{fig:4}(b), see section 3. It consists of a continuous narrow-band laser at frequency $2\omega_0$ pumping a parametric down-converter. After removal of the pump beam, the signal and idler photons are separated. They pass through EOPMs and through narrow-band frequency filters, and are finally detected. We now give a detailed theoretical description of this experiment.

The parametric down-converter produces photon pairs in a frequency-entangled state of the form
\begin{equation} \label{eq:04}
| \Psi \rangle = \int_{-\infty}^{+\infty} \rmd \omega f(\omega) | \omega_0 + \omega \rangle | \omega_0 - \omega \rangle \,,
\end{equation}
where $f(\omega) = r(\omega) \rme^{\rmi\phi(\omega)}$ is a complex function of $\omega$ that characterizes the bandwidth of the signal and idler photons. We have neglected in this expression the linewidth of the pump laser --- whose effect is to make the pump (angular) frequency $2\omega_0$ slightly uncertain.

If one measures the frequencies of Alice and Bob's photon state, one finds perfect correlations: if Alice obtains $\omega_0 + \omega$, Bob obtains $\omega_0 - \omega$. In practice the frequency can only be measured with precision $\Omega_{\mathrm{F}}$  given by the width of the frequency filters used. This leads to the notion of \emph{frequency bin}: all photons whose frequencies are contained in an interval $\bigl[ \omega_{\mathrm{F}} - \frac{\Omega_{\mathrm{F}}}{2}, \omega_{\mathrm{F}} + \frac{\Omega_{\mathrm{F}}}{2} \bigr]$ are grouped into a single frequency bin centred on frequency $\omega_{\mathrm{F}}$.

When a single photon of frequency $\omega$ passes through an EOPM, driven at the RF $\Omega_{\mathrm{RF}}$ (with $\Omega_{\mathrm{RF}} > \Omega_{\mathrm{F}}$) with adjustable amplitude $c$ and phase $\gamma$, it undergoes the unitary transformation
\begin{equation} \label{eq:03}
| \omega \rangle \mapsto \hat{U}(c, \gamma) | \omega \rangle = \sum_{n\in\mathbb{Z}} U_n(c, \gamma) | \omega + n\Omega_{\mathrm{RF}} \rangle \,,
\end{equation}
where
\begin{equation} \label{eq:02}
U_n(c, \gamma) = J_n(c) \rme^{\rmi n(\gamma-\pi/2)}
\end{equation}
and $J_n$ is the $n$th-order Bessel function of the first kind.

Since the action of EOPMs on this state can only change the frequencies by integer multiples of $\Omega_{\mathrm{RF}}$, it is convenient to rewrite the state as
\begin{equation} \label{eq:05}
| \Psi \rangle = \int_{-\Omega_{\mathrm{RF}}/2}^{+\Omega_{\mathrm{RF}}/2} \rmd \omega \sum_{n \in \mathbb{Z}} f(n \Omega_{\mathrm{RF}} + \omega) | \omega_0 + n \Omega_{\mathrm{RF}} + \omega \rangle | \omega_0 - n \Omega_{\mathrm{RF}} - \omega \rangle \,.
\end{equation}
The motivation for re-expressing \eref{eq:04} in this form is that the EOPMs will cause contributions from different values of the index $n$ to interfere, while contributions from different values of the offset parameter $\omega$ will add probabilistically. Indeed, with a sufficiently precise measurement of the frequencies of the photons exiting the EOPMs, we could determine a specific value for $\omega$ and, in retrospect, claim the initial entangled state was
\begin{equation} \label{eq:06}
| \Psi \rangle = \sum_{n\in\mathbb{Z}} f(n \Omega_{\mathrm{RF}} + \omega) | \omega_0 + n \Omega_{\mathrm{RF}} + \omega \rangle | \omega_0 - n \Omega_{\mathrm{RF}} - \omega \rangle \,.
\end{equation}

We can further simplify notation by noting that the actual value of $\omega$ in \eref{eq:06} is of no importance. We therefore drop the parameter $\omega$ from \eref{eq:06} and adopt the discretized version
\begin{equation} \label{eq:07}
| \Psi \rangle = \sum_{n\in\mathbb{Z}} f_n | n \rangle | -n \rangle \,,
\end{equation}
where $| n \rangle$ denotes a photon with a frequency $\omega_{0} + n \Omega_{\mathrm{RF}} + \omega$ for some $\omega \in \bigl[ -\frac{\Omega_{\mathrm{RF}}}{2}, +\frac{\Omega_{\mathrm{RF}}}{2} \bigr]$, and we denote $f_n = r_n \rme^{\rmi\phi_n} = f(n \Omega_{\mathrm{RF}} + \omega)$, $\forall n$.

We will also make the hypothesis that $f_n$ varies slowly with $n$, which is justified if $\Omega_{\mathrm{RF}}$ is very small compared to the frequency range over which $f$ varies. In our experiments the bandwidth of the photon-pair source (the scale over which $f$ changes) is approximately 5~THz, while $\Omega_{\mathrm{RF}}=25$~GHz. This allows us to identify $f_n \approx f_{n+p}$ for small values of $p$, say $-5\leq p\leq +5$ (see next paragraph).

In the experiment schematized in figure~\ref{fig:4}(b), each photon is separately modulated with respective parameters $a,\alpha$ and $b,\beta$. According to \eref{eq:03}, the state \eref{eq:07} is transformed to
\begin{equation} \label{eq:08}
| \Psi \rangle \mapsto \hat{U}(a, \alpha) \otimes \hat{U}(b, \beta) | \Psi \rangle = \sum_{n,d\in\mathbb{Z}} f_{n} c_d(a, \alpha; b, \beta) | n \rangle | -n+d \rangle \,,
\end{equation}
where $c_d(a, \alpha; b, \beta) = \sum_{p\in\mathbb{Z}} U_p(a, \alpha) U_{d-p}(b, \beta)$, and where we use the assumption that $f_{n \pm p} \approx f_{n}$. This is reasonable since $| U_p(c, \gamma) |$ decreases rapidly with $p$ for accessible RF amplitudes: the values of $p$ for which $|U_p|$ is large are limited to approximately $p\in[-5,+5]$.

The joint probability of Alice detecting a photon in the frequency bin $n$ on which frequency filter $\mathrm{F}_{\mathrm{A}}$ is aligned and Bob detecting a photon in bin $-n+d$ on which filter $\mathrm{F}_{\mathrm{B}}$ is aligned is given by
\begin{equation} \label{eq:09}
P_d(a, \alpha; b, \beta; n) = \bigl| \langle n | \langle -n+d | \Psi \rangle \bigr|^2 = | f_n |^2 \bigl| c_d(a, \alpha; b, \beta) \bigr|^2 \,.
\end{equation}

At this stage, we note that the series giving the coincidence probability can be summed by using the Graf addition formula \cite{ref:as:graf} (which we rederive in \ref{sec:graf} in our adopted notation). We have
\begin{equation} \label{eq:10}
\sum_{p \in \mathbb{Z}} U_p(a, \alpha) U_{d-p}(b, \beta) = U_d(C, \Gamma) \,,
\end{equation}
where $C = [ a^2 + b^2 + 2ab \cos(\alpha-\beta) ]^{1/2}$ and $\tan \Gamma = \frac{a \sin \alpha + b \sin \beta}{a \cos \alpha + b \cos \beta}$, and we may always take $C$ to be positive. In terms of these parameters,
\begin{equation} \label{eq:11}
P_d(a, \alpha; b, \beta; n) = | f_n |^2 J_d(C)^2 \,.
\end{equation}
Thus manipulating frequency-entangled photons with EOPMs gives rise to Bessel-type interference patterns, rather than the usual sine and cosine interference patterns in optics when only two modes are present.

Note that \eref{eq:11} implies the normalization
\begin{equation} \label{eq:12}
\sum_d P_d(a, \alpha; b, \beta; n) = | f_n |^2 \sum_d J_d(C)^2 = | f_n |^2
\end{equation}
required by conservation of probability.

Note also that with modulation turned off the photons do not change frequency and we have $P_{d=0}(a = b = 0; n) = | f_n |^2$ and $P_{d\neq0}(a = b = 0; n)=0$, as expected.

Equation \eref{eq:11} shows that in the experiment schematized in figure~\ref{fig:4}(b), the coincidence rate $N^{(2)}_d$ for frequency bins $n$ and $-n+d$ will be given by
\begin{eqnarray} \label{eq:13}
N^{(2)}_d(a, \alpha; b, \beta; n) \nonumber \\
~~~ = J_d \bigl( [ a^2 + b^2 + 2ab \cos(\alpha-\beta) ]^{1/2} \bigr)^2 \times N^{(2)}_{d=0}(a=b=0; n) \,,
\end{eqnarray}
where $N^{(2)}_{d=0}(a=b=0; n)$ is the coincidence rate for frequency bins $n$ and $-n$ when the modulation is off.

\subsection{Bell inequality optimisation \label{sec:bell}}

We now show that the correlations \eref{eq:11} allow the violation of a Clauser-Horne (CH74) inequality \cite{ref:ch74}, specifically the violation of $S\leq2$, where (see \cite{ref:oc10})
\begin{eqnarray} \label{eq:17}
S &=& \bigl[ N_{d=0}^{(2)}(a_0, \alpha_0; b_0, \beta_0; n) + N_{d=0}^{(2)}(a_0, \alpha_0; b_1, \beta_1; n) + N_{d=0}^{(2)}(a_1, \alpha_1; b_0, \beta_0; n) \nonumber \\
&& - N_{d=0}^{(2)}(a_1, \alpha_1; b_1, \beta_1; n) \bigr] / N_{d=0}^{(2)}(a=b=0; n) \,.
\end{eqnarray}

The summation based on the Graf addition formula allows a relatively straightforward determination of the optimal parameters for violating the CH74 inequality. To this end, we first substitute \eref{eq:11} in order to rewrite the CH74 expression as
\begin{equation} \label{eq:18}
S = J_0(C_{00})^2 + J_0(C_{01})^2 + J_0(C_{10})^2 - J_0(C_{11})^2 \,,
\end{equation}
where
\begin{equation} \label{eq:19}
C_{ij} = [ a_i^2 + b_j^2 + 2 a_i b_j \cos(\alpha_i-\beta_j) ]^{1/2}
\end{equation}
and $i, j \in \{0, 1\}$.

The parameters $C_{ij}$ obey constraints imposed by the form of \eref{eq:19}. To see this, we introduce the vectors
\begin{eqnarray} \label{eq:20}
\bi{a}_i &=& ( a_i \cos \alpha_i,\, a_i \sin \alpha_i ) \,, \nonumber \\
\bi{b}_j &=& -( b_j \cos \beta_j,\, b_j \sin \beta_j ) \,.
\end{eqnarray}
In terms of these, $C_{ij} = | \bi{a}_i - \bi{b}_j |$. We may therefore identify the $C_{ij}$ with the lengths of the sides of a quadrilateral defined by the vertex vectors $\bi{a}_0$, $\bi{b}_0$, $\bi{a}_1$ and $\bi{b}_1$. This implies that each of the four $C_{ij}$ is bounded by the sum of the other three. For example,
\begin{eqnarray} \label{eq:21}
C_{11} &=& | \bi{a}_1 - \bi{b}_1 | \nonumber \\
&=& | \bi{a}_1 - \bi{b}_0 + \bi{b}_0 - \bi{a}_0 + \bi{a}_0 - \bi{b}_1 | \nonumber \\
&\leq& | \bi{a}_1 -\bi{b}_0 | + | \bi{b}_0 - \bi{a}_0 | + | \bi{a}_0 - \bi{b}_1 | \nonumber \\
&=& C_{10} + C_{00} + C_{01} \,.
\end{eqnarray}

In this way, we reduce the eight-parameter optimization of \eref{eq:17} to a four-parameter optimization with constraints. There are two possibilities: either the optimum will lie within the parameter domain, or it will lie along one of the boundaries. We quickly rule out the former possibility: in this case, local extrema are found for parameters $C_{ij}$ which are local extrema of $J_0^{\,2}$. Since at the second greatest extremum we have $J_0(x)^2 \approx 0.162$ (for $x \approx 3.832$), no combination of four positive extrema of $J_0^{\,2}$ satisfying the strict inequalities will lead to a violation of the CH74 expression.

A Bell inequality violating optimum, if one exists, must therefore lie along one of the boundaries. In \ref{sec:proof}, we show that the global optimum of \eref{eq:18} lies along the constraint $C_{00} = C_{01} = C_{10} = C_{11}/3$, systematically ruling out any other possibility. The optimum of $C_{00} \approx 0.550$, for which $S \approx 2.389$, corresponds to the RF parameters
\begin{eqnarray} \label{eq:22}
(a_0, \alpha_0) =& (0.275, \theta) &= (b_0, \beta_0) \,, \nonumber \\
(a_1, \alpha_1) =& (0.825, \theta+\pi) &= (b_1, \beta_1) \,.
\end{eqnarray}
Reaching the optimal value requires the use of variable --- but small --- modulation amplitudes and precise phase adjustment.

These optimal parameters should be contrasted with the analysis and experiment reported in \cite{ref:oc10} where the CH74 inequality was violated in a configuration where the amplitudes $a_0=a_1=b_0=b_1$ were all equal, and only the phases $\alpha_0,\alpha_1,\beta_0,\beta_1$ varied. In this configuration, the largest CH74 violation tends to the above, but requires arbitrarily large RF amplitudes.  But when the amplitudes are large, the two-photon interference pattern is much more sensitive to small errors in the RF amplitudes and phases. The above --- optimal --- value of $S$ is attained for rather small values of the RF amplitudes, which makes the experiment much more robust.

For comparison, we note that the maximum value for the expression \eref{eq:17} attainable by quantum theory for systems of dimension 2 is 2.414, which is quite close to the maximum value of 2.389 attainable using EOPMs on frequency-bin entangled photons. However, as the frequency-bin entangled photons belong to a Hilbert space of dimension greater than 2, it may be that the maximum value attainable by some local measurements on the state exceeds this value. We do not know whether this is the case. The algebraic maximum for this expression is 3, which cannot be exceeded by any measurement.

\subsection{Equivalence of two-photon and single-photon interference schemes \label{sec:comparison}}

There is a mathematical correspondence between correlation experiments on maximally entangled states and prepare-and-measure schemes, based on the identity $\langle i | \langle j | U_{\mathrm{A}} \otimes U_{\mathrm{B}} | \Phi^+ \rangle = \langle i | U_{\mathrm{A}} U_{\mathrm{B}}^T | j \rangle / \sqrt{d}$ where $| \Phi^+ \rangle = \sum_{i=1}^d | i \rangle | i \rangle / \sqrt{d}$.  Indeed, the first term in the equality can be interpreted as a measurement on the entangled state $ | \Phi^+ \rangle$ in which Alice projects onto $\langle i | U_{\mathrm{A}}$ and Bob projects onto $\langle j | U_{\mathrm{B}}$, whereas the second term can be interpreted as the preparation by Bob of the state $U_{\mathrm{B}}^T | j \rangle$ and the subsequent projection by Alice onto the state $\langle i | U_{\mathrm{A}}$.

This theoretical correspondence is well established in the context of quantum key distribution, where it is used to demonstrate the equivalence between prepare-and-measure schemes and entanglement-based schemes. However, in general, these two schemes correspond to different experiments because implementing $U_{\mathrm{B}}^T$ is physically different from implementing $U_{\mathrm{B}}$. This is the case, for example, in experiments involving time bins. In our case however, where the transformations $U_{\mathrm{A},\mathrm{B}}$ are realized by EOPMs, $U_{\mathrm{B}}^T = U_{\mathrm{B}}$, in the sense that we have the identity $\langle -n | \hat{U}(a, \alpha) | p \rangle = \langle -p | \hat{U}(a, \alpha) | n \rangle$. The above mathematical identity thus translates to a physical correspondence between transition amplitudes in two-photon and single-photon experiments, with all the RF parameters (amplitudes and phases) kept unchanged.

In practice, this corresponds to the equivalence between the scheme depicted in figure~\ref{fig:4}(b) in which entangled photons are manipulated by EOPMs and that of figure~\ref{fig:4}(a) in which photons belonging to a particular frequency bin are selected by filter F$_{\mathrm{A}}$ and subsequently modulated with parameters $(a, \alpha)$ and $(b, \beta)$. Specifically, the amplitude of detecting photons in frequency bins $-n$ and $n+d$ in experiment \ref{fig:4}(b) is proportional to the amplitude of detecting a photon in frequency bin $n+d$ given that it was prepared in bin $n$ in experiment \ref{fig:4}(a):
\begin{eqnarray} \label{eq:14}
&& \langle -n | \langle n+d | \hat{U}(a, \alpha) \otimes \hat{U}(b, \beta) | \Psi \rangle \nonumber \\
&=& \sum_p f_p \langle n+d | \hat{U}(b, \beta) | -p \rangle \langle -n | \hat{U}(a, \alpha) | p \rangle \nonumber \\
&\simeq& f_n \sum_p \langle n+d | \hat{U}(b, \beta) | -p \rangle \langle -p | \hat{U}(a, \alpha) | n \rangle \nonumber \\
&=& f_n \langle n+d | \hat{U}(b, \beta) \hat{U}(a, \alpha) | n \rangle \,,
\end{eqnarray}
where in line 3 we have invoked the same assumption used to derive \eref{eq:08}, namely that $f_n$ is approximately constant over the experimentally accessible range of frequencies which interfere, and we used the completeness relation $\sum_p | p \rangle \langle p | = 1$.

Thus, in the experiment of figure~\ref{fig:4}(a), if a single photon is initially in bin $n$, the probability that it is detected in bin $n+d$ has the same functional dependence as \eref{eq:11} in the two-photon case. Experimentally, the quantity measured is the photon rate $N^{(1)}_d$, where $d$ denotes the separation between the initial bin $n$ and the final bin $n+d$. Therefore, we have
\begin{eqnarray} \label{eq:15}
N^{(1)}_d(a, \alpha; b, \beta; n) \nonumber \\
\quad = J_d \bigl( [ a^2 + b^2 + 2ab \cos(\alpha-\beta) ]^{1/2} \bigr)^2 \times N^{(1)}_{d=0}(a=b=0; n) \,,
\end{eqnarray}
where $N^{(1)}_{d=0}(a=b=0; n)$ is the photon rate for frequency bin $n$ when the modulation is off.

Moreover, if bin $n$ initially contains a large number of photons, then in the experiment of figure~\ref{fig:4}(a) the optical power $N^{(\mathrm{class})}_d$ measured in bin $n+d$, if the initial light beam is prepared in bin $n$, is given by
\begin{eqnarray} \label{eq:16}
N^{(\mathrm{class})}_d(a, \alpha; b, \beta; n) \nonumber \\
\quad = J_d \bigl( [ a^2 + b^2 + 2ab \cos(\alpha-\beta) ]^{1/2} \bigr)^2 \times N^{(\mathrm{class})}_{d=0}(a=b=0; n) \,,
\end{eqnarray}
where $N^{(\mathrm{class})}_{d=0}(a=b=0; n)$ is the optical power for frequency bin $n$ when the modulation is off.

In addition to the correspondence to the entanglement-based scheme, experiments based on figure~\ref{fig:4}(a) are of interest in themselves: when used with a single-photon source, this scheme allows the realization of quantum communication protocols such as quantum key distribution, see \cite{ref:bm07}.

\subsection{A note on phase accumulation during propagation}

If a photon of frequency $\omega$ propagates a distance $L$, its state is transformed according to
\begin{equation} \label{eq:23}
| \omega \rangle \mapsto \rme^{\rmi k(\omega)L} | \omega \rangle \simeq \rme^{\rmi(\beta_0 + \beta_1 (\omega -\omega_0)+\cdots)L} | \omega \rangle \,,
\end{equation}
where we have developed the wave number $k$ in series in $\omega -\omega_0$.

When considering frequency coding of information, the zeroth-order term $\rme^{\rmi\beta_0 L}$ is an overall phase, and has no physical influence. The first-order term $\rme^{\rmi \beta_1 (\omega -\omega_0)L}$ can be absorbed into the phase $\gamma$ of the RF field applied to the EOPM, see \eref{eq:02}. In all calculations as well as the analysis of the experiments, we absorb the phases $\rme^{\rmi \beta_1 (\omega -\omega_0)L}$ into the phases of the RF fields.

The fact that phases accumulated during propagation can be absorbed into the phases of the RF fields underlies the inherent high stability of experiments using frequency bins. Indeed, stability of our experimental setup requires that $\beta_1 \Omega_{\mathrm{RF}} L \ll 1$ (since the only frequencies that interfere are those separated by small multiples of $\Omega_{\mathrm{RF}}$). This should be contrasted with interference experiments in the spatial domain, where one is sensitive to the phase $\beta_0 L$. Approximate equality of the phase and group velocities implies that $\beta_0 \simeq \beta_1 \omega_0$. Since $\Omega_{\mathrm{RF}}/\omega_0 \simeq 10^{-4}$, our experiments are less sensitive to changes in fibre lengths by a factor of roughly $10^{-4}$. This implies that in laboratory experiments, no stabilization is required. In field experiments, however, where propagation distances are tens of kilometres, the local RF oscillators must be synchronized, as was done for instance in \cite{ref:md02}.

Note that the higher-order terms in \eref{eq:23} due to frequency dispersion cannot be absorbed in the phase $\gamma$ of the RF field. In this work, we neglect dispersion effects, but it may degrade the quality of interference in certain cases --- i.e. dispersion management may be needed for long-distance and broad-spectrum applications.

\section{Experimental setups \label{sec:setup}}

\begin{figure}[htbp]
\centering
\includegraphics[scale=.9]{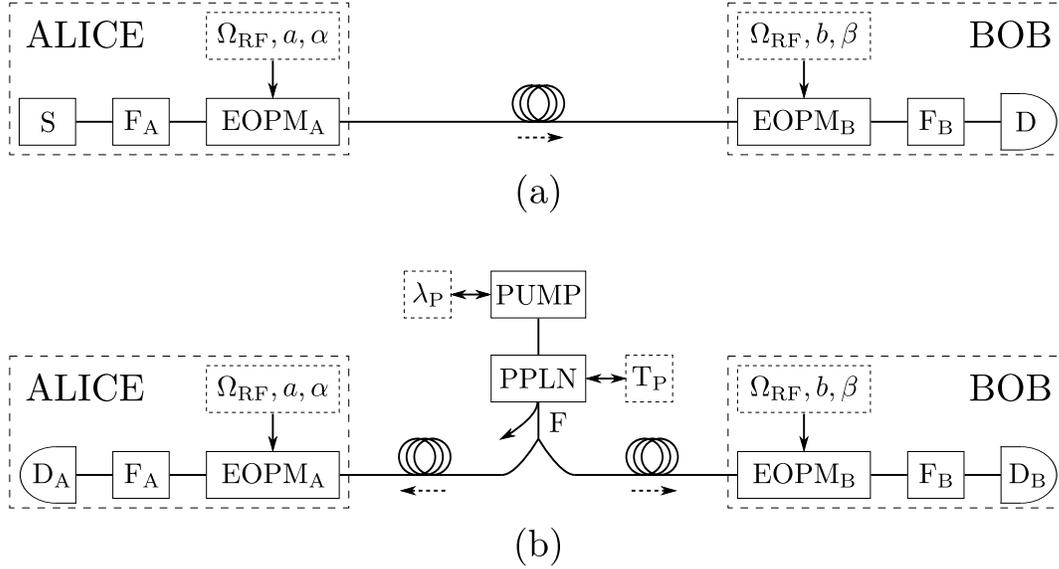}
\caption{Schematic representation of the experimental setups. (a) Classical optics and single-photon experiments. The source S can be a narrow-band laser, a broadband incoherent source or a broadband incoherent source attenuated to reach the single-photon regime. The detector D is either a classical photodiode or, in the single-photon regime, an APD operated in the Geiger mode. (b) Two-photon experiment. The source consists of a PUMP laser (whose wavelength $\lambda_{\mathrm{P}}$ is stabilized) that enters a PPLN waveguide (whose temperature $T_{\mathrm{P}}$ is stabilized) and is then removed by a drop filter F. Detectors $\mathrm{D}_{\mathrm{A}}$ and $\mathrm{D}_{\mathrm{B}}$ are SSPDs. In both panels, band-pass filters $\mathrm{F}_{\mathrm{A}}$ and $\mathrm{F}_{\mathrm{B}}$ select photons belonging to a specific frequency bin. The phase of the light is modulated by $\mathrm{EOPM}_{\mathrm{A}}$ and $\mathrm{EOPM}_{\mathrm{B}}$, driven by a signal at frequency $\Omega_{\mathrm{RF}}$ with respective parameters $a, \alpha$ and $b, \beta$, determined by the RF system depicted in figure~\ref{fig:5}.}
\label{fig:4}
\end{figure}

In this section we describe the experimental implementations of the setups already introduced and illustrated in figures~\ref{fig:4}(a) and (b), and point out the critical requirements for high-visibility experiments.

\subsection{Classical optics and one-photon experiments}

We present here in total three versions of the experiment schematized in figure~\ref{fig:4}(a), differing mainly in the optical source S employed. In the first version, S is a classical source consisting of a coherent polarized narrow-band laser. In this case neither filtration nor polarization management is required, since all the photons emitted already belong to a given frequency bin, with their polarization aligned with the active axes of the modulators. In the second version, the source, also classical, consists of a non-coherent non-polarized broadband light source. In this case, the light beam must first pass through the filter $\mathrm{F}_{\mathrm{A}}$ and a polarizer before being sent through the setup. This corresponds more strictly to the entangled-photon case, where signal and idler photons have a broad spectrum. In the third version (the one-photon experiment), an ideal broadband single-photon source is approximated by attenuating the broadband light source until it contains, on average, much fewer than one photon in each frequency bin within each detection time window. We report the results for all three sources.

In the experiment, filter $\mathrm{F}_{\mathrm{A}}$ is a home-made fibre Bragg grating used in reflection with a circulator. It selects photons belonging to the frequency bin $\bigl[ \omega_{\mathrm{F}} - \frac{\Omega_{\mathrm{F}}}{2}, \omega_{\mathrm{F}} + \frac{\Omega_{\mathrm{F}}}{2} \bigr]$ centred on $\omega_{\mathrm{F}} = \omega_0$ and restricted to a 3~dB width $\frac{\Omega_{\mathrm{F}}}{2\pi} \approx 3~\mathrm{GHz}$. Photons are consecutively guided through $\mathrm{EOPM}_{\mathrm{A}}$ and $\mathrm{EOPM}_{\mathrm{B}}$ (EOSPACE) driven by an RF signal at $\frac{\Omega_{\mathrm{RF}}}{2\pi} = 25~\mathrm{GHz}$ with adjustable amplitude and phase $a, \alpha$ and $b, \beta$, respectively (controlled by the setup depicted in figure~\ref{fig:5}). At 12.5~GHz, the isolation of $\mathrm{F}_{\mathrm{A}}$ is better than 30~dB, ensuring that frequency bins are well isolated from each other. Photons selected by filter $\mathrm{F}_{\mathrm{B}}$ (which has the same characteristics as $\mathrm{F}_{\mathrm{A}}$) are detected. Where classical sources are used, the optical power $N^{(\mathrm{class})}$ is measured by a photodetector D. In the single-photon regime, the single-photon rate $N^{(1)}$ is measured with an APD (id Quantique) operated in gated mode (repetition rate = 100~kHz, detection window size = 2.5~ns, efficiency $\approx$ 10\% and dark count rate $\approx 3 \times 10^{-6}~\mathrm{ns}^{-1}$).

\subsection{Two-photon experiment}

In the experiment schematized in figure~\ref{fig:4}(b), a narrow-band (2~MHz bandwidth) continuous pump laser (Sacher Lasertechnik) stabilized at a wavelength $\lambda_\mathrm{P}$ emits 3~mW of power into a 3~cm long PPLN waveguide (HC Photonics). Phase matching is achieved by controlling the waveguide's temperature $T_{\mathrm{P}}$. It is then possible to efficiently generate frequency-entangled photon pairs centred on a specific frequency, corresponding in our case to the wavelength $2\lambda_{\mathrm{P}} = 2\pi c / \omega_0 = 1547.743~\mathrm{nm}$.

At the output of the PPLN, a drop filter F rejects the pump with more than 60~dB isolation. The pairs are separated (with 50\% probability) with a 3~dB coupler, such that each photon in every entangled pair is sent through an independent EOPM, $\mathrm{EOPM}_{\mathrm{A}}$ or $\mathrm{EOPM}_{\mathrm{B}}$. Polarization-maintaining fibre components ensure that the photons' polarizations are aligned with the active axes of the modulators. Filters $\mathrm{F}_{\mathrm{A}}$ and $\mathrm{F}_{\mathrm{B}}$ select photons belonging to the frequency bin centred on $\omega_0$. Photons are finally detected by SSPDs $\mathrm{D}_{\mathrm{A}}$ and $\mathrm{D}_{\mathrm{B}}$ (Scontel) cooled to 1.7~K and operated in the continuous mode (efficiency $\approx$ 5\%; dark count rate $\approx$ 30~Hz).

A time-to-digital converter performs a coincidence measurement. Specifically, it registers the arrival times $t_{\mathrm{A}}$ and $t_{\mathrm{B}}$ of photons A and B and records the number of coincident detection events as a function of $t_{\mathrm{B}} - t_{\mathrm{A}}$. The coincidence rate $N^{(2)}$ is extracted from the histogram by summing all contributions for which the difference in arrival times $t_{\mathrm{B}} - t_{\mathrm{A}}$ is contained in a given time interval of width 0.6~ns.

\subsection{Requirements for high-visibility experiments}

In order to produce high-visibility Bessel interference patterns, our architecture must fulfil some critical requirements, particularly in the two-photon case.

\begin{figure}[htbp]
\centering
\includegraphics[scale=.9]{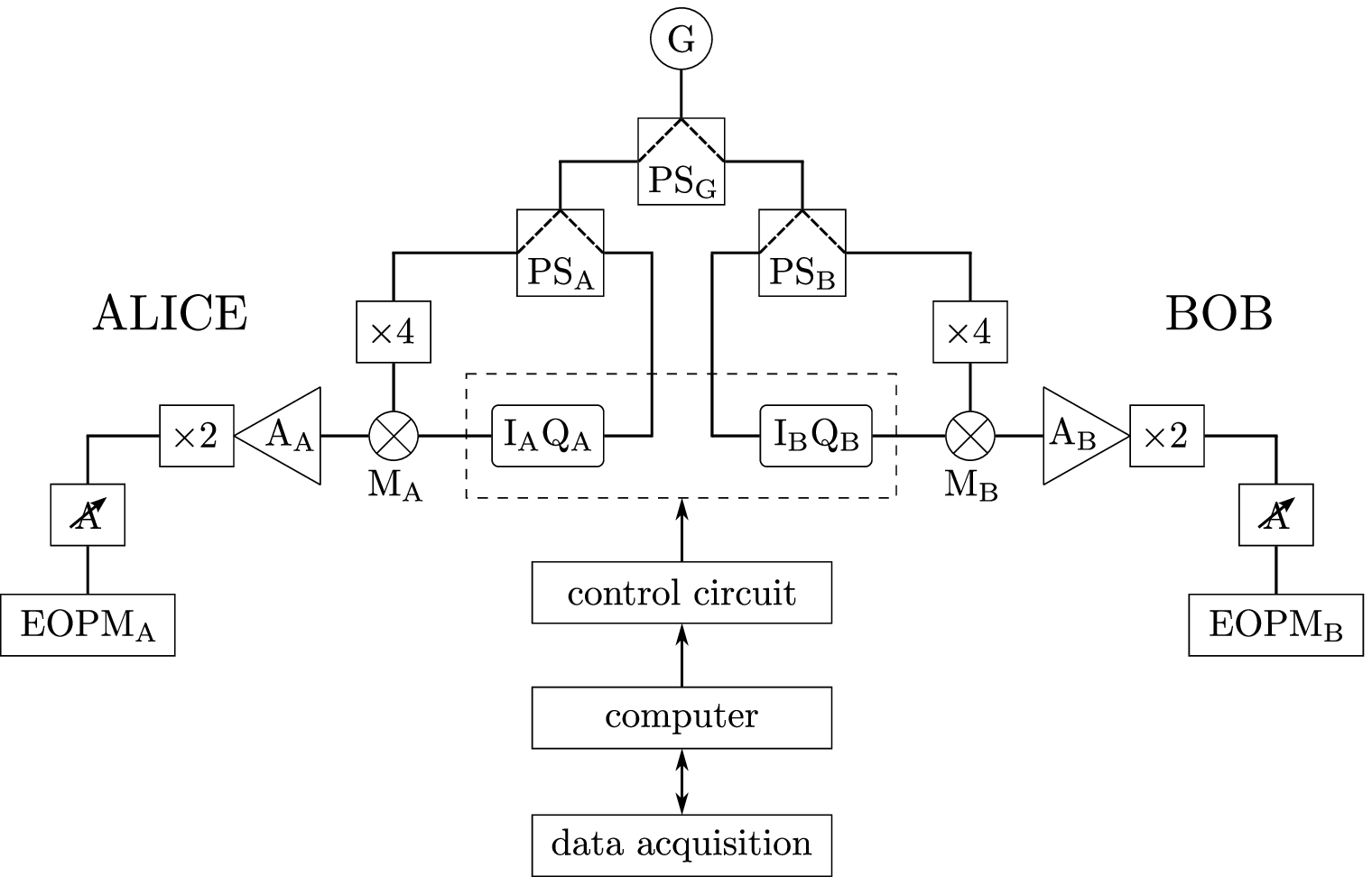}
\caption{The RF system used to drive $\mathrm{EOPM}_{\mathrm{A}}$ and $\mathrm{EOPM}_{\mathrm{B}}$ in all experiments. A 2.5~GHz signal of an RF generator G is separated by power splitters $\mathrm{PS}_\mathrm{G}$, $\mathrm{PS}_\mathrm{A}$ and $\mathrm{PS}_\mathrm{B}$. The phase of both Alice's and Bob's signals is controlled by I\&Q modulation. Using frequency multipliers $\times 4$ and $\times 2$, signal multipliers $\mathrm{M}_\mathrm{A}$ and $\mathrm{M}_\mathrm{B}$ and amplifiers $\mathrm{A}_\mathrm{A}$ and $\mathrm{A}_\mathrm{B}$, strong $25$~GHz signals are obtained. Their amplitude is controlled by variable attenuators.  Accurate control of the parameters $a, \alpha$ and $b, \beta$ is thus obtained.}
\label{fig:5}
\end{figure}

First, in order to obtain precise reproducible interference in the frequency domain, high-resolution phase and amplitude control of the RF signals must be achieved. We use the RF architecture presented in figure~\ref{fig:5}. This architecture is based on RF translation and uses cheap off-the-shelf RF components. It could therefore be adapted for field experiments in which Alice and Bob are separated by a large distance.

In this RF architecture, an RF generator outputs a 2.5~GHz, 10~dBm signal. Its power is equally shared by splitter $\mathrm{PS}_\mathrm{G}$ between Alice and Bob. The two parts of the setup are equivalent and made independent with isolators. Alice's and Bob's signals are split by $\mathrm{PS}_{\mathrm{A}}$ and $\mathrm{PS}_{\mathrm{B}}$, respectively. A first part of the signal is subjected to I\&Q modulation: a given combination of attenuation allows the precise imposition of an RF phase. A digital-to-analogue converter controls with high precision the I\&Q voltages $I_{\mathrm{A}}, Q_{\mathrm{A}}$ and $I_{\mathrm{B}}, Q_{\mathrm{B}}$, allowing fine selection of $(I, Q)$ pairs. In this way, phases $\alpha$ and $\beta$ can be shifted from 0 to $2\pi$ while the amplitude remains constant. Since the phase shift introduced at 2.5~GHz is altered by frequency translation (see below), a mapping is performed at 25~GHz in order to identify the correct $(I, Q)$ couples. This permits distortion correction, resulting in a highly accurate control of the phase of the RF signal with a precision better than $10^{-2}~\mathrm{rad}$. The second part of the signal is sent through a multiplier and the 10~GHz signal obtained is multiplied, with fixed phase, by the phase-shifted signal to obtain a 12.5~GHz signal with adjustable phase. Frequency filters control the purity of the signal. The signal is sent through an amplifier and a frequency doubler, which amplifies the signal and converts it into 25~GHz. Finally, the signal is applied to $\mathrm{EOPM}_{\mathrm{A}}$ or $\mathrm{EOPM}_{\mathrm{B}}$ with an adjustable amplitude $a = \pi V_\pi^{-1} R^{1/2} P_{\mathrm{A}}^{1/2}$ or $b = \pi V_\pi^{-1} R^{1/2} P_{\mathrm{B}}^{1/2}$, where the half-wave voltage $V_\pi \approx 2.9~\mathrm{V}$, the internal resistance $R = 50~\Omega$ and power $P_{\mathrm{A}}$ or $P_{\mathrm{B}}$ is mechanically adjusted with a variable attenuator. Ten per cent of the signal is sent to an RF power meter allowing an adjustment of $P_{\mathrm{A}}$ and $P_{\mathrm{B}}$ with $10^{-2}~\mathrm{dB}$ precision.

Second, in order to achieve a high and stable SNR in the two-photon case, fine control of the pump laser wavelength is compulsory. Each down-converted photon must belong to a frequency bin whose width is set by the 3~GHz band-pass filters $\mathrm{F}_{\mathrm{A}}$ and $\mathrm{F}_{\mathrm{B}}$. The pump wavelength must thus be accurately set at the centre of a frequency bin. To this end, 10\% of the power of the pump laser is sent to a wavelength meter (Exfo) that generates an electric signal proportional to the difference between the measured wavelength and a reference. A proportional-integral-derivative loop generates an error signal which feeds the piezoelectric transducer of the external cavity of the laser, stabilizing the wavelength at $\lambda_{\mathrm{P}} = \left( 773.8715 \pm 0.0002 \right)~\mathrm{nm}$. The degeneracy frequency $\omega_0$ is thus controlled with a precision of around $\pm 0.04~\mathrm{GHz}$.

Third, in order to have a high and stable pair production rate, one must also optimize the parametric down-conversion of the PPLN source. To achieve a fine conversion wavelength control and good efficiency, the PPLN crystal is seated on a Peltier cell for accurate temperature stability. Precise measurements with a tunable C-band laser in the second-harmonic regime show that after several hours, a stable optimized conversion wavelength is reached with about 1~pm precision.

Achievement of the latter two conditions allows us to detect (using low-noise SSPDs) coincidences at a rate of approximately 20~Hz and with a SNR of approximately $2 \times 10^3$ (without RF signal applied to the EOPMs).

Together with the accurate control of the RF parameters, these improvements lead to raw and net visibilities of $(99.17\pm0.11)\%$ and $(99.76\pm0.11)\%$, respectively. This should be compared to the $98\%$ net visibilities reported in \cite{ref:oc10}.

Note that compared to our previous work \cite{ref:oc10}, control of the RF phases is now automated and the accuracy much improved. The detectors in all experiments can be interfaced to a computer, enabling both selection of RF phases $\alpha$ and $\beta$ and measurement data acquisition. Rates $N^{(2)}$, $N^{(1)}$ and $N^{(\mathrm{class})}$ are thus measured during automatically adjustable times in automatically adjustable RF configurations. Note that replacing mechanically variable attenuators by remotely controlled components would allow full automatization of the system. Note also that for long-distance experiments it is possible to use two synchronized RF generators instead of one, as shown in \cite{ref:md02}.

\section{Experimental results \label{sec:results}}

\subsection{Equivalence of interference schemes}

We first evaluate the rates $N^{(2)}$, $N^{(1)}$ and $N^{(\mathrm{class})}$. As discussed in section \ref{sec:comparison}, these should have identical dependence on the RF parameters, see \eref{eq:13}, \eref{eq:15} and \eref{eq:16}. To experimentally test this equivalence, we chose $d=0$, and the parameter $C = [ a^2 + b^2 + 2ab \cos(\alpha-\beta) ]^{1/2}$ is varied by scanning one of the phases $\alpha$ or $\beta$ with $a=b$ fixed. This procedure permits easy evaluation of the interference visibility, the value of which is critical for the performance of the system.

We define the \emph{raw} and \emph{net} visibilities as follows:
\begin{eqnarray} \label{eq:24}
V_{\mathrm{raw}} &=& \frac{N_{\max} - N_{\min}}{N_{\max} + N_{\min}} \,, \\
\label{eq:25}
V_{\mathrm{net}} &=& \frac{( N_{\max} - N_{\mathrm{noise}} ) - ( N_{\min} - N_{\mathrm{noise}} )}{( N_{\max} - N_{\mathrm{noise}} ) + ( N_{\min} - N_{\mathrm{noise}} )} \,,
\end{eqnarray}
where $N$ denotes either $N^{(2)}$, $N^{(1)}$ or $N^{(\mathrm{class})}$, and $N_{\mathrm{noise}}$ in \eref{eq:25} represents the noise due to detector imperfections (e.g. dark counts) which can be measured independently.

The maximal rate $N_{\max}$ is obtained in principle for $C^*=0$, such that $J_0 (C^*) ^2 = 1$. For $a=b$, this is achieved with a phase difference $\alpha - \beta = \pi$. The minimal rate $N_{\min}$ is obtained for any of the positive roots $\{C^*_{i}\}$ of $J_0$, for which $J_0 (C^*_{i}) ^2 = 0,\,\forall i$. The visibility therefore attains a theoretical maximum value of 1. If $a = b$, the first root $C^*_{1}$ is attainable at sufficiently high RF powers ($a = b \gtrsim 1.2$) with the phase difference $\alpha - \beta = \arccos (C_{1}^{*2} / 2a^2 - 1)$.

\begin{figure}[htbp]
\centering
\includegraphics[scale=.9]{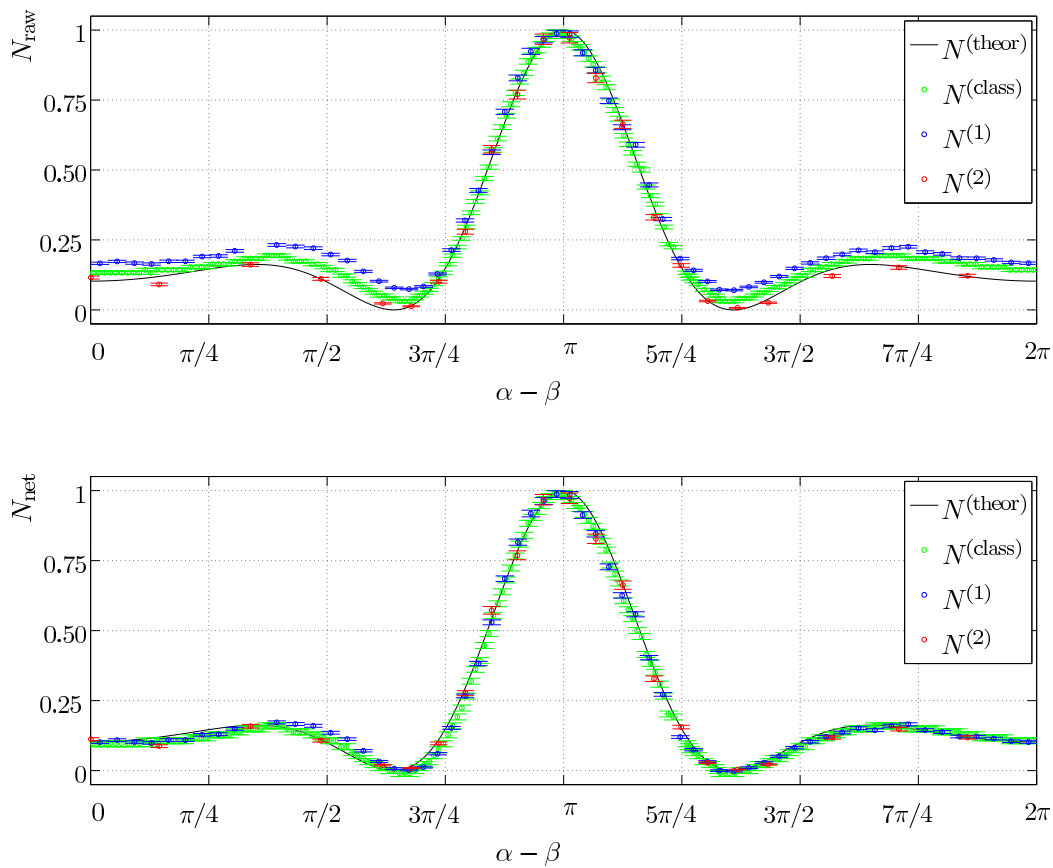}
\caption{Bessel interference pattern in the frequency domain. The interference described by \eref{eq:13}, \eref{eq:15} and \eref{eq:16} is plotted for $d = 0$ and $a = b \approx 2.25$ as a function of the phase difference $\alpha - \beta$. The raw (with detector noise) and net (detector noise subtracted) data are shown, respectively, on the top and bottom panels. On both plots, the theoretical curve is included in black, the green data points are the normalized light intensity $N^{(\mathrm{class})}$ (obtained using the tunable laser source and the setup of figure~\ref{fig:4}(a)), the blue data points correspond to the normalized photon rate $N^{(1)}$, i.e. the single-photon interference pattern (obtained using the broadband source, attenuated to the single-photon regime, and the setup of figure~\ref{fig:4}(a)), and the red data points correspond to the normalized coincidence rate $N^{(2)}$, i.e. the two-photon interference pattern (obtained using the PPLN source and the setup of figure~\ref{fig:4}(b)). The error bars are statistical. Note that the classical measurements $N^{(\mathrm{class})}$ are made with a relatively noisy photodiode, and the raw visibility for the green data points in the top panel is therefore less than that reported in table \ref{tab:1}, where a low-noise photodiode was used. Note also that when detector noise is subtracted (bottom panel) the different experimental data points superpose exactly, demonstrating the equivalence of the interference schemes, but deviate slightly from the theoretical curve. We attribute this to small errors in the calibration of I\&Q parameters, so that the actual phases $\alpha$ and $\beta$ deviate slightly from their theoretical value.}
\label{fig:6}
\end{figure}

Our results are summarized in figure~\ref{fig:6} and in table~\ref{tab:1}. Figure~\ref{fig:6} is a plot of $J_d \bigl( [ a^2 + b^2 + 2ab \cos(\alpha-\beta) ]^{1/2} \bigr)^2$ as a function of $\alpha - \beta$ for $d = 0$ and $a = b \approx 2.25$. The experimentally derived rates $N^{(2)}$, $N^{(1)}$ and $N^{(\mathrm{class})}$ --- normalized by $N^{(2)}$, $N^{(1)}$ and $N^{(\mathrm{class})}$ with modulation turned off --- are superposed to the theoretical curve. The close agreement of the experimental data taken with the different experimental schemes validates the predictions of \eref{eq:13}, \eref{eq:15} and \eref{eq:16} and thereby demonstrates the equivalence of the two-photon, one-photon and classical optics experiments of figure~\ref{fig:4}.

Note that, as mentioned in section~\ref{sec:setup}, an uncontrolled phase shift arises due to propagation times in the optical fibres. The experimental results are therefore horizontally shifted in order to obtain the best possible agreement with the theoretical curve. This is the only parameter that is adjusted to fit the data.

Table~\ref{tab:1} gives the values of the visibilities $V_{\mathrm{raw}}$ and $V_{\mathrm{net}}$ extracted from curves such as those reported in figure~\ref{fig:6}. In the classical case a very low noise detector was used to measure maximal and minimal rates, which is why $V_{\mathrm{raw}} \approx V_{\mathrm{net}}$. Both the laser source and the broadband source were used. In the single-photon case an APD with a relatively high dark count rate was used, which explains the low value of the raw visibility. In the entangled-photon case, the high value of $V_{\mathrm{raw}}$ is due to the quality of the SSPDs, which are much less noisy than APDs. Maximal and minimal rates were measured for several minutes in order to obtain good statistical precision on the visibilities.

\begin{table}[htbp]
\caption{Experimentally measured visibilities.}
\label{tab:1}
\vspace{3mm}
\centering
\begin{tabular}{c c c c c}
\hline
& Classical expt & Classical expt & One-photon expt & Two-photon expt \\
& (laser source) & (broadband source) & & \\ \hline
$V_{\mathrm{raw}}$ & $(99.79\pm0.01)\%$ & $(99.41\pm0.12)\%$ & $(87.25\pm0.38)\%$ & $(99.17\pm0.11)\%$ \\
$V_{\mathrm{net}}$ & $(99.79\pm0.01)\%$ & $(99.41\pm0.12)\%$ & $(99.27\pm0.43)\%$ & $(99.76\pm0.11)\%$ \\ \hline
\end{tabular}
\end{table}

The values of $V_{\mathrm{raw}}$ depend strongly on the noise inherent in the detectors and thus vary greatly across the different cases. By contrast, the values of $V_{\mathrm{net}}$ are all almost equal and notably high. This agreement once again confirms the equivalence between the different experiments. It also allows us to separate the contributions to visibility degradation due to the experimental setup from those inherent to the detectors (detector noise and dark counts) and sources (in particular noise due to multiple photon events) used. The main contributions from the setup itself are imperfect frequency-bin isolation, imperfect polarization management and imperfect control of the RF parameters.

\subsection{Bell inequality violation}

As a second main result, we evaluate the experimental violation of the CH74 inequality introduced in section~\ref{sec:bell}. Scans of both phases $\alpha$ and $\beta$ at the amplitudes $(a_0 , b_0)$, $(a_0 , b_1)$, $(a_1 , b_0)$, $(a_1 , b_1)$ given in \eref{eq:22} enable precise selections of the phases $\alpha_0 , \beta_0 , \alpha_1 , \beta_1$ given in \eref{eq:22} for which the violation will be largest. Measurements are then realized for phases and amplitudes optimizing the violation.

The experimental results, listed in table~\ref{tab:2}, agree with the theoretical predictions up to the statistical errors and imply that the CH74 inequality is violated by more than 18 standard deviations.

\begin{table}[htbp]
\caption{Bell inequality violation results. $A_i B_j$, $i, j \in \{0, 1\}$ is the notational shorthand for $(a_i, \alpha_i; b_j, \beta_j)$. The column “Theory" gives the values that should be obtained at the point \eref{eq:22} for which the CH74 violation is maximal. Note that the experimental values agree with theoretical predictions up to statistical errors.}
\label{tab:2}
\vspace{3mm}
\centering
\begin{tabular}{c c c c}
\hline
& Theory & Expt $N^{(2)}_{\mathrm{raw}}$ & Expt $N^{(2)}_{\mathrm{net}}$ \\
& & (with noise) & (noise subtracted) \\ \hline
$A_0 B_0$ & 0.857  & $0.862 \pm 0.006$ & $0.861 \pm 0.006$  \\
$A_0 B_1$ & 0.857  & $0.863 \pm 0.006$ & $0.862 \pm 0.006$  \\
$A_1 B_0$ & 0.857  & $0.854 \pm 0.006$ & $0.853 \pm 0.006$  \\
$A_1 B_1$ & 0.182  & $0.190 \pm 0.003$ & $0.186 \pm 0.003$  \\ \hline
$S$       & 2.389  & $2.389 \pm 0.021$ & $2.391 \pm 0.021$  \\ \hline
\end{tabular}
\end{table}

We should, however, mention that this result does not provide a decisive test of local causality as we have not closed either the detection or locality loopholes, and because the CH74 inequality, as we have applied it, requires additional assumptions (see the discussion in \cite{ref:oc10,ref:ch74}). Nevertheless, these results show that our present approach allows the study of quantum correlations of frequency-entangled photons, and could in principle, i.e. if experimental imperfections (mainly losses and detector inefficiencies) were small, be adapted to permit a decisive test of local causality.

\section{Conclusion \label{sec:conclusion}}

In summary, building on our earlier work \cite{ref:oc10}, we have further developed the theory underlying the manipulation of frequency-bin entangled photons with EOPMs, and demonstrated that this could be reliably realized experimentally. At the theoretical level, our main results are an analytic simplification of the expression giving the coincidence probabilities, and the demonstration of the equivalence of prepare-and-measure schemes with two-photon schemes. In our experiment, frequency-entangled photons are produced by parametric down-conversion with a frequency-stabilized pump laser and a temperature-stabilized PPLN waveguide. They are made to interfere in the frequency domain through EOPMs whose driving signals are controlled by a dedicated RF architecture built entirely from off-the-shelf components. Interference patterns are detected with SSPDs via optical frequency filtering. The interference patterns are accurately controlled, exhibit high visibilities and allow the violation of a Bell inequality by 18 standard deviations.

The strengths of our method are
\begin{itemize}
\item the use of optical, electro-optic and RF components that are commercially available and allow easy interconnection and remote control;
\item the use of optical components that allow good polarization management, frequency-bin isolation and stability;
\item the use of an RF system that allows stability, independence, easy calibration and precise adjustment of parameters;
\item overall reproducibility and robustness allowing day-long experiments with no measurable drift or decrease in performance.
\end{itemize}
Our system can be easily adapted to field experiments, as it is robust and allows full automatization and long-distance synchronization. We are therefore confident that we will be able to extend our results in the near future to long-distance quantum communication experiments and to perform tasks such as long-distance Bell inequality violation or quantum key distribution. These future experimental works would be supported by the theoretical analysis introduced here.

\ack

We acknowledge support from the Belgian Science Policy under project Photonics@be (IAP contract P6/10), from the French Agence Nationale de la Recherche under the project “High bit-rate and versatile quantum-secured networks" (HQNET contract 032-05) and from the Conseil R\'egional de Franche-Comt\'e. IM acknowledges support from the Conseil G\'en\'eral du Doubs. EW acknowledges support from the Belgian Fonds pour la Formation \`a la Recherche dans l'Industrie et l'Agriculture. Finally, we thank Fran\c{c}ois Leo for his assistance in stabilizing the frequency of our pump laser and Samuel Moec for his help with PID temperature control.

\appendix

\section{The Graf addition formula \label{sec:graf}}

The matrix elements $U_n(c, \gamma) = J_n(c) \rme^{\rmi n (\gamma - \pi/2)}$ are the Fourier components of the periodic function $\rme^{- \rmi c \cos (\phi - \gamma)}$, so we have
\begin{equation} \label{eq:a1}
\rme^{-\rmi c \cos(\phi - \gamma)} = \sum_n U_n(c, \gamma) \rme^{-\rmi n \phi}
\end{equation}
(the Jacobi-Anger expansion). Since adding two sine waves with the same period yields another sine wave, we may write
\begin{equation} \label{eq:a2}
\rme^{- \rmi a \cos(\phi - \alpha)} \rme^{- \rmi b \cos(\phi - \beta)} = \rme^{-\rmi C \cos(\phi - \Gamma)} \,.
\end{equation}
We extract an expression for $C$ and $\Gamma$ by applying the identity $\cos(\phi - \gamma) = \cos \phi \cos \gamma + \sin \phi \sin \gamma$ and comparing terms in $\sin \phi$ and $\cos \phi$:
\begin{eqnarray} \label{eq:a3}
a \cos \alpha + b \cos \beta &=& C \cos \Gamma \,, \\
\label{eq:a4} a \sin \alpha + b \sin \beta &=& C \sin \Gamma \,,
\end{eqnarray}
from which we obtain
\begin{equation} \label{eq:a5}
C^2 = a^2 + b^2 + 2 a b \cos(\alpha - \beta)
\end{equation}
and
\begin{equation} \label{eq:a6}
\tan \Gamma = \frac{a \sin\alpha + b \sin\beta}{a \cos\alpha + b \cos\beta} \,.
\end{equation}
Note that we may impose $C \geq 0$, which fully determines $\Gamma$. Equations~\eref{eq:a1} and \eref{eq:a2} and the convolution theorem then imply
\begin{equation} \label{eq:a7}
\sum_n U_n(a, \alpha) U_{d-n}(b, \beta) = U_d(C, \Gamma) \,.
\end{equation}

Intuitively, \eref{eq:a2} and \eref{eq:a7} express that two phase modulators used in series, driven by sinusoidal RF signals of the same frequency, have the same action as a single phase modulator. Equation \eref{eq:a7} is a slight generalisation of the Graf addition formula as given in \cite{ref:as:graf}, expressed in the notation we have adopted.

\section{Optimization of CH74 \label{sec:proof}}

We begin with the boundary $C_{11} = C_{00} + C_{01} + C_{10}$. Considering the partial derivatives of the expression
\begin{equation} \label{eq:b1}
S = J_0(C_{00})^2 + J_0(C_{01})^2 + J_0(C_{10})^2 - J_0(C_{00} + C_{01} + C_{10})^2 \,,
\end{equation}
we find that the optimum must satisfy
\begin{equation} \label{eq:b2}
(J_0^{\,2})'(C_{00}) = (J_0^{\,2})'(C_{01}) = (J_0^{\,2})'(C_{10}) = (J_0^{\,2})'(C_{00} + C_{01} + C_{10}) \,,
\end{equation}
where $(J_0^{\,2})'$ is the derivative of $J_0^{\,2}$. Due to the symmetry in \eref{eq:b1}, we may impose $C_{00} \leq C_{01} \leq C_{10}$ without loss of generality. Since $J_0^{\,2} \leq 1$, a violation of $S \leq 2$ will clearly require $C_{00} \leq x_{2/3} \approx 0.878$, such that $J_0(C_{00})^2 \geq 2/3$, and $C_{01} \leq x_{1/2} \approx 1.126$, where $J_0(C_{01})^2 \geq 1/2$. Because there is no overlap in the images $(J_0^{\,2})'([0, x_{2/3}])$ and $(J_0^{\,2})'(]x_{2/3}, x_{1/2}])$, \eref{eq:b2} further imposes $C_{01} \leq x_{2/3}$. Finally, \eref{eq:b2} together with injectivity of $(J_0^{\,2})'$ on the domain $[0, x_{2/3}]$ impose $C_{00} = C_{01}$, so we require $C_{00} = C_{01} \leq x_{2/3}$.

We first assume $C_{00} = C_{01} = C_{10} \leq x_{2/3}$ and $C_{11} = 3 C_{00}$, reducing the problem to a one-parameter optimization of
\begin{equation} \label{eq:b3}
S = 3 J_0(C_{00})^2 - J_0(3 C_{00})^2 \,.
\end{equation}
Numerically, we find an optimal value of $S_{\mathrm{optimized}} \approx 2.389$ for $C_{00} \approx 0.550$.

We now show that this is the global optimum. The alternative would be to take some $C_{10} > x_1$. In this case, it will not be possible to satisfy \eref{eq:b2} unless $C_{10} \gtrsim 1.291$, and we have no chance of surpassing the optimum given above unless $J_0(C_{10})^2 \gtrsim 0.389$, which translates to $C_{10} \lesssim 1.293$. Via \eref{eq:b2}, this in turn imposes $C_{00} = C_{01} \gtrsim 0.876$, for which $2 J_0(C_{00})^2 \lesssim 1.634$. However, $J_0(C_{10})^2 \lesssim 0.390$ for $1.291 \lesssim C_{10} \lesssim 1.293$, which guarantees we will not surpass the optimum of \eref{eq:b3}.

Finally, we check that there is not a better optimum along one of the three other possible boundaries. Using the symmetry of \eref{eq:18}, without loss of generality we consider the boundary $C_{10} = C_{00} + C_{01} + C_{11}$. In this case, and assuming that the function $J_0^{\,2}$ is symmetric, the analogue of \eref{eq:b1} is
\begin{equation} \label{eq:b4}
S = J_0(C_{00})^2 + J_0(C_{01})^2 + J_0(C_{10})^2 - J_0(C_{00} + C_{01} - C_{10})^2
\end{equation}
and we find that an optimum along this boundary must satisfy
\begin{equation} \label{eq:b5}
(J_0^{\,2})'(C_{00}) = (J_0^{\,2})'(C_{01}) = (J_0^{\,2})'(-C_{10}) = (J_0^{\,2})'(C_{00} + C_{01} - C_{10}) \,.
\end{equation}
Equations~\eref{eq:b4} and \eref{eq:b5} are identical to \eref{eq:b1} and \eref{eq:b2}, except with the substitution $C_{10} \mapsto -C_{10}$. In this case, we still require $C_{00} = C_{01} \leq x_{2/3}$. This combined with \eref{eq:b5} would impose $- C_{10} \lesssim -2.405$, where $J_0^{\,2} \lesssim 0.162$, again guaranteeing that we will not surpass the optimum found in the preceding case.

\section*{References}

\bibliographystyle{iopart-num}
\bibliography{fb}

\end{document}